\documentclass[
twocolumn,
reprint,
superscriptaddress,
nofootinbib,
nobibnotes,
amsmath,
amssymb,
aps,
longbibliography,
prb,
showkeys
]{revtex4-2}

\usepackage[latin1]{inputenc}
\usepackage{amsmath}
\usepackage{amsfonts}
\usepackage{hyperref}
\usepackage{amssymb}
\usepackage{graphicx}
\usepackage{hyperref}
\usepackage{colortbl}
\usepackage{subfigure}
\usepackage{amsthm}
\usepackage{thmtools}
\usepackage[normalem]{ulem}
\usepackage{dsfont}
\usepackage{circuitikz}
\usepackage{cases}

\usepackage[T1]{fontenc}
\usepackage{mathptmx}
\usepackage{bbold}
\usepackage{scalerel}
\usepackage{xcolor}

\usepackage{mathtools}

\numberwithin{equation}{section}

\newcommand{\hyref}[1]{\hyperref[#1]{\ref{#1}}}
\newcommand{\dd}{\mathrm{d}}

\newcommand{\orange}[1]

\renewcommand{\thesection}{\arabic{section}}
\renewcommand{\thesubsection}{.\arabic{subsection}}

\newcommand{\thenewsubsection}{\thesection.\arabic{subsection}}
\newcommand{\thenewsubsubsection}{\thesection\thesubsection.\arabic{subsubsection}}

\makeatletter
\def\@hangfrom@section#1#2#3{\@hangfrom{#1#2}#3}
\def\@hangfroms@section#1#2{#1#2}
\makeatother

\usepackage{titlesec}

\titleformat{\section} 
{\fontsize{12}{14}\bfseries} 
{\thesection} 
{1em} 
{\centering} 
[] 

\titleformat{\subsection} 
{\fontsize{10}{12}\bfseries} 
{\thenewsubsection} 
{2em} 
{\centering} 
[] 

\titleformat{\subsubsection} 
{\fontsize{10}{12}\itshape} 
{\thenewsubsubsection} 
{2em} 
{\centering} 
[] 
\begin{document}

\title{Nonlinear iontronic signal processing with neuromorphic Spike Rate-Dependent Plasticity}

\author{T. M. Kamsma}
\thanks{These two authors contributed equally to this work}
\affiliation{Institute for Theoretical Physics, Utrecht University, Princetonplein 5, 3584 CC Utrecht, The Netherlands}
\affiliation{Mathematical Institute, Utrecht University, Budapestlaan 6, 3584 CD Utrecht, The Netherlands}
\author{Y. Gu}
\thanks{These two authors contributed equally to this work}
\affiliation{School of Aeronautics and Institute of Extreme Mechanics, Northwestern Polytechnical University, Xi'an 710072, China}
\author{D. Shi}
\affiliation{School of Physical Science and Technology, Northwestern Polytechnical University, Xi'an 710129, China}
\author{C. Spitoni}
\affiliation{Mathematical Institute, Utrecht University, Budapestlaan 6, 3584 CD Utrecht, The Netherlands}
\author{M. Dijkstra}
\affiliation{Soft Condensed Matter \& Biophysics, Debye Institute for Nanomaterials Science,
Utrecht University, Princetonplein 1, 3584 CC Utrecht, The Netherlands}
\author{R. van Roij}
\affiliation{Institute for Theoretical Physics, Utrecht University, Princetonplein 5, 3584 CC Utrecht, The Netherlands}
\author{Y. Xie}
\affiliation{School of Aeronautics and Institute of Extreme Mechanics, Northwestern Polytechnical University, Xi'an 710072, China}

\date{\today}

\begin{abstract}
We present an integrated iontronic memristor circuit that reproduces biologically inspired Spike Rate-Dependent Plasticity (SRDP) and functions as a physical nonlinear frequency kernel, which we demonstrate can be used to classify natural auditory data. The fluidic circuit integrates two parallel memristive membranes containing short and long conical memristive channels with opposite orientations, giving rise to heterogeneous internal timescales and different potentiation responses. As a result, the circuit exhibits a nonlinear frequency response in which low-frequency inputs decrease the overall conductance, whereas higher-frequency inputs increase it, thereby emulating biological SRDP. Our experimental measurements are inspired by and consistent with predictions of a theoretical model. We demonstrate the functionality of the device by separating encoded sound signals from different insects that cannot be linearly separated. By unifying theoretical predictions with experimental realisation of coupled iontronic memristors, this work moves beyond isolated components and demonstrates how heterogeneous iontronic dynamics can unlock nonlinear time-series processing capabilities, essential for future iontronic neuromorphic computing.

\end{abstract}

\maketitle

\section{Introduction}
Neuromorphic engineering seeks to replicate the brain's computational capabilities by developing hardware that emulates the adaptive dynamics of biological neurons and synapses. Iontronic devices, which rely on ionic and molecular transport in aqueous media, form an intriguing platform for this purpose due to their inherent biomimetic medium, directly mirroring the ion- and molecule-mediated signal transmission of biological neural systems. Despite a plethora of recent progress \cite{Fan2025EmergingComputation,Lv2025AdvancementsDesign,SuwenLaw2025RecentComputing,Wang2026NeuromorphicConductors}, current iontronic platforms face several challenges, including scaling beyond single-device fabrication \cite{EdriFraiman2025TowardIntegration} and identifying clear pathways toward practical applications \cite{Fan2025EmergingComputation}. The (real-time) processing of temporal data through so-called reservoir computing has emerged as a promising direction for iontronic devices \cite{Kamsma2024Brain-inspiredNanochannels,Armendarez2024Brain-InspiredPlasticity,Portillo2025NeuromorphicDiodes,Kamsma2025EchoSubstrate,Li2025BiologicallyComputing,Portillo2026ReservoirMemristors}, possibly offering considerable improvements in energy-efficiency over existing substrates \cite{Kamsma2026Energy-efficientCircuits} and direct coupling to biological signals \cite{Kamsma2025EchoSubstrate}. In these reservoir computing protocols, fixed iontronic recurrent neural network circuits map dynamic inputs to high-dimensional physical states within the circuit, which can be used to straightforwardly classify the original input via a simple trainable readout function \cite{Cucchi2022Hands-onImplementation}. Standard (physical) readout functions are linear crossbar arrays \cite{Liu2025Resistance-RestorableChip,VanDeBurgt2018OrganicComputing,Kazemzadeh2025AllArray,Xu2025Angstrom-Scale-ChannelComputing,Zhang2022AdaptiveConductor,Hu2023AnComputing,VanDeBurgt2017AComputing, Zolfagharinejad2025AnalogueComputing}, therefore nonlinear responses within the dynamic circuit are of significant importance \cite{Zolfagharinejad2025AnalogueComputing}. Additionally, to move toward such iontronic reservoir computing circuits, multiple (memristive) devices need to be integrated and respond to shared dynamic inputs, which remains experimentally challenging. In this work we present an integrated iontronic circuit that represents simultaneous advances in both challenges of integrating multiple devices and extracting nonlinear dynamics, thereby enabling nonlinear audio processing as real-world time series data showcase.

\begin{figure*}[ht!]
\centering
\includegraphics[width=1\textwidth]{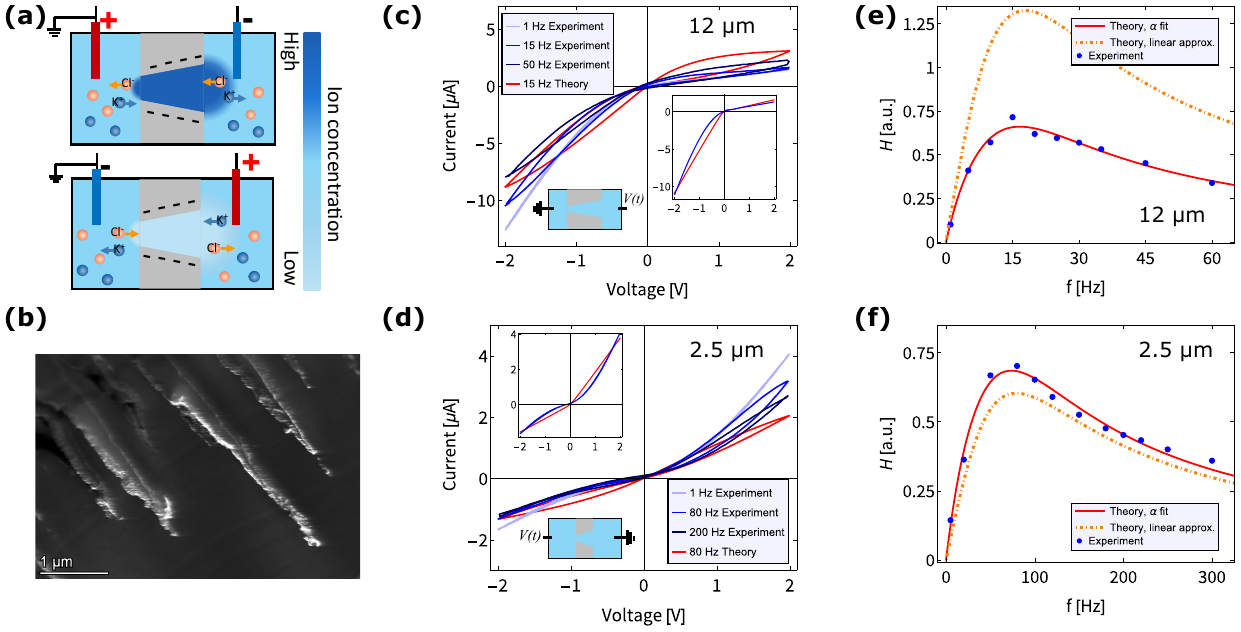}
\caption{\textbf{(a)} Schematic of ion concentration polarisation in a conical channel with negative surface charge under applied voltages. \textbf{(b)} Transmission electron microscopy image of a $\sim100$ nm thick cross-sectional slice (cut parallel with the channels) from the 12 $\mu$m thick membrane. \textbf{(c,d)} Current-voltage hysteresis loops for the long (c) and short (d) channels with grounded tip and base, respectively, at various frequencies, from experiments (blue) and the theoretical model (red). Insets show the steady-state currents measured experimentally (blue) and predicted by the theoretical model for conical channels \cite{Kamsma2023UnveilingIontronics} (see text). \textbf{(e,f)} Frequency-dependence of the dimensionless enclosed current-voltage area $H$ (see text) in the high conductance lobe for the long (e) and short (f) channels, measured experimentally (blue) and predicted theoretically (orange and red). The orange curves show the first-order relation for the conductance $g_{i,0}h_{i,\infty}=g_{i,0}(1+\alpha_i V/V_{\mathrm{r}})$ fitted to the experimentally found steady-state conductance, with $V_{\mathrm{r}}=1$\,V. The agreement with the overall functional form is good, which becomes evident when $\alpha$ is treated as a fit parameter, shown by the red curves.}
\label{fig:Fig1}
\end{figure*}

Here, we integrate two types of iontronic conical channel memristors \cite{Wang2012TransmembraneTransport,Kamsma2023IontronicMemristors,Ramirez2023SynapticalMemristors,Ramirez2024MemristivePores} and extract a non-monotonic, and thus nonlinear, frequency response that emulates the hallmark synaptic feature of Spike Rate-Dependent Plasticity (SRDP), exhibited by various biological neurons, whereby low-frequency neural activity weakens synaptic connection strength, while high-frequency firing results in potentiation \cite{Rick1996FrequencyRat,Xu2008GABABSynapses,Kumar2011Frequency-dependentPlasticity}. This nonlinear biologically inspired frequency response enables the classification of sound signals from bees and two of their predators, dragonflies and hornets, data which are not linearly separable based on their characteristic frequencies. The memristors within the fluidic SRDP circuit are of different channel lengths and orientations. The different channel lengths give rise to distinct retention timescales \cite{Kamsma2023IontronicMemristors,Cervera2024ModelingDiodes,Kamsma2024Brain-inspiredNanochannels,Zhang2024GeometricallySystems}, allowing the combined circuit to respond to both low- and high-frequency inputs. Additionally, the selected channel orientations give rise to opposing conductance responses, such that the long channels act as an inhibiting component, whereas the short channels function as a potentiating component. These design features enable a compact integrated fluidic memristor circuit that transitions between synaptic depression and potentiation as a function of input frequency, thereby reproducing characteristic neuronal SRDP.

The characteristics of the individual channel membranes, as well as the resulting SRDP behaviour, are predicted and characterised by a theoretical model. The key features of the underlying physics are well captured by our theory, although achieving full quantitative agreement remains challenging due to geometric polydispersity of the channels within the membranes. Due to this theoretical characterisation the emerging SRDP properties of this device are well understood. Moreover, the theoretical model can help identify potential applications and guide future experiments exploring other nonlinear frequency response properties.

To demonstrate the functionality of our device we carry out an elementary time series analysis of real-world data. We encode sound signals from bees and two of their predators, dragonflies and hornets, as voltage pulses. These are applied to the device, which converts the pulse trains to distinct conductances. Notably, the predators' signals are not linearly separable from the bees based on sound signal frequencies and require a nonlinear kernel before linear classification, e.g.\ via a crossbar array, is possible. Due to its reliable nonlinear frequency response, our device provides such a conversion. Therefore, our compact integrated iontronic circuit functions as a synaptic nonlinear frequency kernel for temporal data, providing a straightforward pathway to end-to-end integration with standard (linear) readout functions such as (ionic) crossbar arrays \cite{Liu2025Resistance-RestorableChip,VanDeBurgt2018OrganicComputing,Kazemzadeh2025AllArray,Xu2025Angstrom-Scale-ChannelComputing,Zhang2022AdaptiveConductor,Hu2023AnComputing,VanDeBurgt2017AComputing,Zolfagharinejad2025AnalogueComputing}.

\section{Individual channel characterisation}
For SRDP with iontronic memristors, components are required that respond differently to low- and high-frequency inputs and can exhibit both decreasing and increasing conductance responses. Here we show how these features can straightforwardly be implemented by varying the channel orientations and lengths of conical channels, which are well known to exhibit ion concentration polarisation (ICP), i.e.\ the accumulation or depletion of ions within the channel, depending on the polarity of the applied voltage \cite{wei1997current, Boon2022Pressure-sensitiveGeometry, white2008ion, vlassiouk2009biosensing,cheng2007rectified, siwy2006ion, bush2020chemical, siwy2002rectification,fulinski2005transport,siwy2005asymmetric}. As schematically depicted in Fig.~\ref{fig:Fig1}(a), for conical channels of a negative surface charge  a negative potential at the base compared to the tip results in ion accumulation, thereby increasing the conductance of the channel. Ion depletion occurs for a positive potential at the base compared to the tip, thereby decreasing the conductance of the channel. Therefore, by grounding either the tip or the base side of the channel, opposite conductance responses are achieved when applying a voltage $V$ at the base or tip side respectively. Additionally, this ICP is transient and the timescales over which occurs provide conical channels with a volatile conductance memory \cite{Wang2012TransmembraneTransport,Kamsma2023IontronicMemristors,Ramirez2023SynapticalMemristors,Ramirez2024MemristivePores}. This conductance memory time depends on the length of the channel with longer channels having a longer response time \cite{Kamsma2023IontronicMemristors}.

We fabricated membranes with  thicknesses of 12 $\mu$m and 2.5 $\mu$m containing  conical channels using an asymmetric latent-track etching technique \cite{apel_diode-like_2001,siwy_fabrication_2002,zhou_nanofluidic_2024}. Further details are provided in Materials and Methods. The channel lengths are therefore determined by the membrane thickness, resulting in long and short conical channels that are 12 $\mu$m and 2.5 $\mu$m in length, respectively. Using transmission electron microscopy on a $\sim100$ nm thick cross-sectional slice (cut parallel to the channel axis) from the 12 $\mu$m thick membrane reveals the conical nanochannels shown in Fig.~\ref{fig:Fig1}(b). As visible in Fig.~\ref{fig:Fig1}(b), the membranes in our experiments feature a polydisperse set of conical channels. For our present purposes, the key properties are the orientation dependence of ICP and the length dependence of the conductance retention time, rather than the precise conductance contribution of each individual channel. Consequently, for the theoretical analysis we consider each membrane as a single representative conical channel and scale the conductance of channels in membrane $i=\text{s,l}$, representing short and long channels, respectively, by a suitable zero-field  conductance $g_{i,0}$. 

For a single channel, labeled by index $i$, the time-dependent conductance $g_i(t)$ due to a time-dependent voltage $V(t)$ is predicted to trace a voltage-dependent steady-state conductance $g_{i,0}h_{i,\infty}(V(t))$ \cite{Kamsma2023IontronicMemristors,Kamsma2023UnveilingIontronics}, described by
\begin{align}\label{eq:gODE}
\frac{\dd g_{i}(t)}{\dd t}=\frac{g_{i,0}h_{i,\infty}(V(t))-g_i(t)}{\tau_i},
\end{align}
determined by a characteristic conductance memory timescale $\tau_i$, an equilibrium conductance $g_{i,0}$, and a dimensionless voltage-dependent steady-state conductance $h_{i,\infty}(V)$. Here, $V(t)$ is the potential applied at the base for the 12 $\mu$m membrane (with grounded tip side), and applied at the tip for the 2.5 $\mu$m membrane (with grounded base side). The steady-state conductances $h_{i,\infty}(V(t))$ were calculated using the theoretical model of Ref.~\cite{Kamsma2023UnveilingIontronics} (more details in Materials and Methods). Numerical values for $\tau_i$ and $g_{i,0}$ will be discussed later. In Figs.~\ref{fig:Fig1}(c) and (d) we show the experimentally measured currents upon applying a periodic triangular voltage input, with a grounded tip for the long channels in (c) and a grounded base for the shorter channels in (d), at various frequencies in blue and the theoretical prediction obtained by numerically solving Eq.~(\ref{eq:gODE}) for one representative frequency, shown in red. Clear hysteresis loops are found, confirming the channels are iontronic memristors, with reasonable agreement between theory and experiment.

The relaxation timescale $\tau_i$ dictates a diffusionlike conductance memory retention time of order ms, predicted to be dependent on the channel length $L_i$ \cite{Kamsma2023IontronicMemristors}. For pristine single channels $\tau_i\propto L_i^2$ \cite{Kamsma2023IontronicMemristors,Kamsma2024Brain-inspiredNanochannels,Cervera2024ModelingDiodes}, while the membranes here feature a polydisperse set of channels with varying geometries. Additionally, we note that, on top of the diffusionlike volatile memory timescale $\tau_i$, there is a range of additional conductance memory timescales observed in conical channels \cite{Jouveshomme2025MultipleMem-spectrometry,Rivera-Sierra2025RelaxationApplications}. Also in this work the short channel membranes exhibit such a longer timescale, possibly due to the potential presence of underetched angstrom-scale channels \cite{lu_angstrom-scale_2025} with longer relaxation times \cite{Shi2023UltralowMemristor,Xu2025Angstrom-Scale-ChannelComputing}. Nevertheless, we will show that a clear length dependence is still found. To model the membranes, $\tau_i$ is determined from a theoretical analysis of the experimentally observed frequency-dependence of their current-voltage hysteresis loops.

\begin{figure*}[ht!]
\centering
\includegraphics[width=1\textwidth]{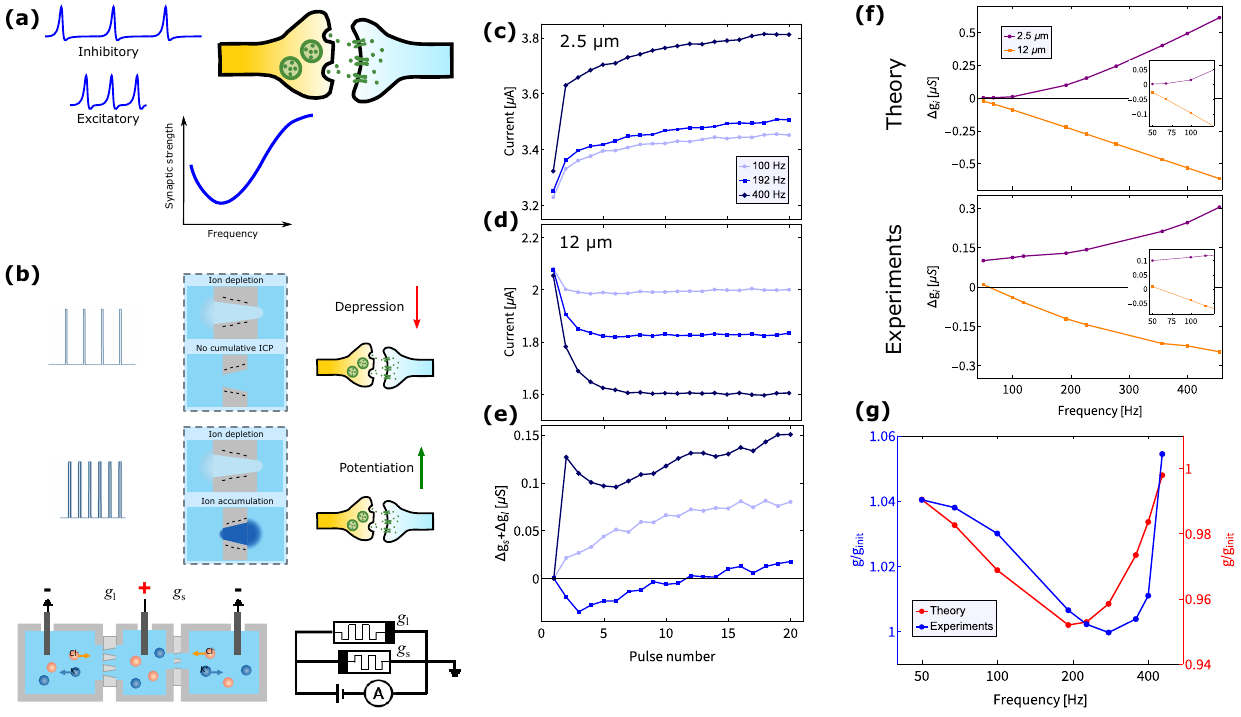}
\caption{\textbf{(a)} Illustration of biological spike rate-dependent plasticity (SRDP), where the synaptic strength exhibits a minimum at intermediate frequencies. \textbf{(b)} Schematic of the driving mechanism behind iontronic SRDP and (bottom) the integrated iontronic SRDP circuit, consisting of two conical channel membranes with thickness $L_{\text{s}}=2.5\;\mu\text{m}$ and $L_{\text{l}}=12.0\;\mu\text{m}$ and opposing channel orientations. \textbf{(c,d,e)} Measured current responses of the individual membranes of short (c) and long (d) channels to an applied train of voltage pulses for three different frequencies, which combine to produce three distinct net conductance changes shown in (e). \textbf{(f)} Theoretical predictions (top) and experimental measurements (bottom) of the relative conductance changes (when converged) in the long (orange) and short (purple) channel membranes as a function of frequency. Insets highlight the (negligible) conductance change at low frequencies for short channels (purple). \textbf{(g)} Total conductance as a function of frequency as predicted by theory (red) and measured experimentally (blue), showing a clear analogy with biological SRDP.}
\label{fig:Fig2}
\end{figure*}

As shown in Figs.~\ref{fig:Fig1}(c) and (d), hysteresis loops produced at different frequencies enclose different surface areas. A general device-agnostic expression for the (dimensionless) enclosed area $H$ of a current-voltage hysteresis loop under a sinusoidal voltage of frequency $f$ was recently derived for a broad class of Simple Volatile Memristors (SVMs) \cite{Kamsma2024ACircuits}, which are well described by Eq.~(\ref{eq:gODE}). As the loop component corresponding to the high-conductance regime is easier to analyse experimentally, and a triangular, rather than a sinusoidal, voltage waveform is applied here, we rederive the expression of $H$ for the high conductance loop to first order, i.e.\ we assume a linear voltage-dependent steady-state conductance $h_{i,\infty}(V)=1+\alpha_i V/V_{\mathrm{r}}+\mathcal{O}(V^2)$, with $V_{\mathrm{r}}=1$ V a characteristic voltage scale. For a triangular potential with period $1/f$, this yields 
\begin{equation}\label{eq:H}
\begin{split}
	H(f)=4\left|\alpha\right|&\left(\frac{V_0}{V_{\mathrm{r}}}\right)^3\left[f\tau-8(f\tau)^2\tanh{\left(\frac{1}{4f\tau}\right)}\right.\\
	&\left.+32(f\tau)^3\left(1-\frac{1}{\cosh{\left(\frac{1}{4f\tau}\right)}}\right)\right],
\end{split}
\end{equation}
with $V_0/V_{\mathrm{r}}$ the (dimensionless) amplitude of the triangular voltage potential. Although this expression appears to be quite different from the first-order form for the sinusoidal voltage potential \cite{Kamsma2024ACircuits}, the actual function form is strikingly similar. We stress that Eq.~(\ref{eq:H}), as well as the expression for a sinusoidal voltage \cite{Kamsma2024ACircuits}, represents a general, device-agnostic relation that applies to all SVMs.

The dimensionless quantity $H$ is related to the experimentally measured (dimensional) enclosed area $X$ according to $X=g_{i,0}V_{\mathrm{r}}^2H$. This relation can be used to extract the characteristic timescale $\tau$, which according to Eq.~(\ref{eq:H}) satisfies the relation $2\pi f_{\text{max}}\tau\approx0.93$, where $f_{\text{max}}$ denotes the frequency at which $H(f)$ is maximal. Here we note that this differs from the relation $2\pi f_{\text{max}}\tau=1$ that is predicted for sinusoidal voltages \cite{Kamsma2024ACircuits}. The scaling of $X$ (and thereby $H$) involves two parameters, $\tau$ and $g_{i,0}\alpha_i$, neither of which affects the functional form, as the timescale $\tau$ only rescales the (horizontal) frequency axis while the parameter $g_{i,0}\alpha_i$ only rescales the (vertical) area axis. In Figs.~\ref{fig:Fig1}(e) and (f) we show the experimentally obtained values of $H$ in blue for the long and short channels, respectively. From this data, we extract the characteristic timescales of $\tau_{\mathrm{l}}=8.9$ ms for the long channels and $\tau_{\mathrm{s}}=2.0$ ms for the short channels. Through a first-order fit $g_{i,0}h_{i,\infty}=g_{i,0}(1+\alpha_i V/V_{\mathrm{r}})+\mathcal{O}(V^2)$, fitted to the steady-state I-V curves shown in the insets of Figs.~\ref{fig:Fig1}(c) and (d), we find estimates for $g_{i,0}$ and $\alpha_i$. For clarity we reiterate that the theoretical red graphs in the insets of Figs.~\ref{fig:Fig1}(c) and (d) are not these first-order expansions, but the full theoretical model \cite{Kamsma2023UnveilingIontronics} (more details in Materials and Methods). Using the found values of $g_{i,0}\alpha_i$, we plot the corresponding theoretical predictions for $H$ in orange in Figs.~\ref{fig:Fig1}(e) and (f), with long-channel parameters $g_{\mathrm{l},0}=2.68\;\mu\text{S}$ and $\alpha_{\mathrm{l}}=-0.50$, and short-channel parameters $g_{\mathrm{s},0}=0.79\;\mu\text{S}$ and $\alpha_{\mathrm{s}}=0.23$. For the long channels, the predicted magnitude of $H$ differs from experiment by a factor $\sim2$, whereas for the short channels, we find a reasonable quantitative agreement. Moreover, the qualitative functional dependence of $H$ on frequency shows excellent agreement, which becomes clear when we rescale the vertical axis of the theoretical predictions for $H$ by fitting $g_{i,0}\alpha_i$, as shown in red in Figs.~\ref{fig:Fig1}(e) and (f), yielding $\alpha_{\mathrm{l}}=-0.40$ and $\alpha_{\mathrm{s}}=0.42$ (assuming the same $g_{i,0}$ as before). These results confirm the predicted frequency-dependence of $H$ given by Eq.\ (\ref{eq:H}) experimentally, providing initial evidence for this general predicted relation of any SVM, albeit with rescaled axes at this stage.

\section{Iontronic Spike Rate-Dependent Plasticity}
Neurons can exhibit a nonlinear synaptic response to spike-train frequencies, typically referred to as spike rate-dependent plasticity (SRDP). Coarsely speaking, this entails that low-frequency neural activity weakens synaptic connection strength, whereas high-frequency firing strengthens it \cite{Rick1996FrequencyRat,Xu2008GABABSynapses,Kumar2011Frequency-dependentPlasticity}. This behavior provides a nonlinear relation between spike frequency and synaptic strength, as schematically depicted in Fig.~\ref{fig:Fig2}(a).

The distinct memory timescales of the long and short channels along with their opposite voltage responses arising from opposite channel orientations with respect to ground, as demonstrated in Fig.~\ref{fig:Fig1}, provide the key ingredients required to experimentally realize iontronic SRDP. To construct an iontronic fluidic device that exhibits SRDP we integrate two membranes of small thickness $L_{\text{s}}=2.5\;\mu\text{m}$ and large thickness $L_{\text{l}}=12.0\;\mu\text{m}$ connected in parallel as schematically depicted in the bottom of Fig.~\ref{fig:Fig2}(b), where the long channel membrane is grounded at the tip, while the short membrane is grounded at the base. The two parallel current pathways through the two membranes provide a total conductance $g(t)$ that is given by the sum of individual conductances,
\begin{align}
g(t)=g_{\text{s}}(t)+g_{\text{l}}(t).
\end{align}
Due to their opposite orientations, such that the long channels are tip-grounded while the short channels are base-grounded, a steady positive voltage causes ion accumulation and an increased conduction in the short channels while it causes ion depletion and a decreased conduction in the long channels. 

\begin{figure*}[ht!]
\centering
\includegraphics[width=1\textwidth]{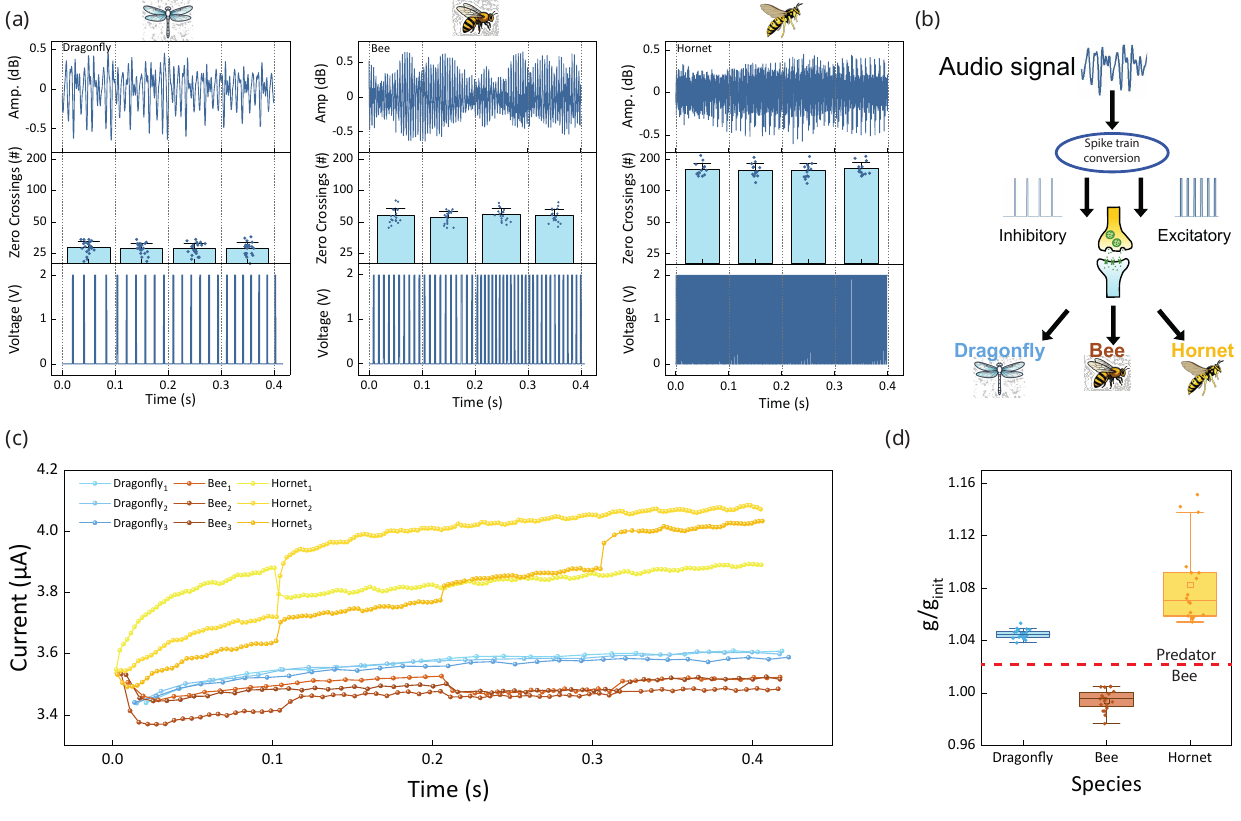}
\caption{\textbf{(a)} Examples of original audio signals and their conversion to spike trains based on the zero crossings of the audio signal for a dragonfly (left), bee (middle), and hornet (right). Dragonflies feature a typical lower pitch, resulting in low-frequency signals, whereas bees produce intermediate-frequency and hornets high-frequency spike trains. \textbf{(b)} Schematic illustration of the inhibitory-excitatory nonlinear synaptic frequency response that enables classification of the insect sound signals. \textbf{(c)} Total circuit current in response to several (pulse-train converted) insect sounds signals, showing that each signal reliably produces a distinguishable current trace per insect. \textbf{(d)} Results for the final conductance from 20 measurements per insect with variance, showing that the different insect signals are not only reliably separated, but also nonlinearly transformed. The intermediate-frequency bee signals become now clearly separated from the lower- and higher-frequency predator signals.}
\label{fig:Fig3}
\end{figure*}

The distinct memory retention timescales enable a frequency regime of trains of positive voltage pulses where the long channels exhibit cumulative ion depletion, while the short channels do not exhibit cumulative ion accumulation as they essentially reset in between pulses. Therefore, in this low-frequency regime the total conductance decreases with frequency. By contrast, in the high-frequency regime the cumulative ion accumulation in the short channels is tuned such that it dominates the cumulative ion depletion in the long channels. Therefore the total conductance increases with frequency in this regime, as schematically illustrated in Fig.\ref{fig:Fig2}(b). We will now demonstrate how this theoretical prediction is experimentally realised.

Spike trains of square-wave voltage pulses with amplitude 2\,V and duration $\Delta t_{\text{d}}=2$ ms are applied with varying interval times $\Delta t_{\text{int}}$, corresponding to frequencies in the range 0 Hz - 455 Hz. The resulting experimentally measured current of the individual short and long membranes driven by spike trains with frequencies 100, 192, and 400 Hz are presented in Figs.~\ref{fig:Fig2}(c) and (d), respectively. As expected on the basis of the steady-state response to positive voltages, for all three frequencies the current (and hence the conductance) increases with the number of applied pulses for the short channels in (c), while they decrease for the longer channels in (d). Notably, the two lower frequencies 100 Hz and 192 Hz converge to similar conductance values for the short channels (Fig.~\ref{fig:Fig2}(c)), whereas the current magnitude, and thus the conductance, of the long-channel membrane (Fig.~\ref{fig:Fig2}(d)) converges to three distinct decreased values for all frequencies. As a result, the frequency-dependence of the total conductance in the two-membrane circuit is non-monotonic, the change in total conductance $\Delta g_{\mathrm{s}}+\Delta g_{\mathrm{l}}$ compared to the total conductance $g_{\text{init}}$ during the first pulse exhibits a nonlinear ordering. This is shown in Fig.~\ref{fig:Fig2}(e), where the experimental time trace at intermediate frequency (192 Hz) shows a smaller conductance change than at 100 Hz and 400 Hz.

These experimental findings agree with our theoretical predictions shown in Fig.~\ref{fig:Fig2}(f), where we compare the theoretically predicted changes in conductance $\Delta g_i$, with $i=\mathrm{s,l}$, of the individual channels (top) to the experimentally found conductance changes (bottom) for the full range of measured frequencies. At low frequencies $f\ll1/\tau_{\text{s}}$ the inter-pulse interval exceeds $\tau_{\text{s}}$, allowing the short channels to reset in between the pulses. The longer channels do respond cumulatively to lower frequencies, since they retain their altered conductance during the longer intervals between pulses and cumulative ion depletion occurs. Hence the overall conductance $g_{\text{s}}(t)+g_{\text{l}}(t)$ decreases over time during low-frequency voltage trains. This effect is visible in the insets of Fig.~\ref{fig:Fig2}(f), where we see that the short channels (purple) change little in conductance for lower frequencies, while the conductance of the longer channels does decrease. Interestingly, for the short channels there seems to always be a small initial low-frequency increase in conductance, resulting in a positive conductance offset. This offset might be caused by the presence of the aforementioned underetched angstrom-scale channels that potentiate for a longer term under an applied voltage \cite{Shi2023UltralowMemristor,Xu2025Angstrom-Scale-ChannelComputing}, which is an effect that is not included in our current theory. At high frequencies $f\sim1/\tau_{\text{s}}$ the shorter channels do display cumulative ICP and therefore increase in conductance. By precisely engineering the individual membrane conductances in the memristive circuit, we ensured that the increase in shorter channels' conductance exceeds the long channels' conductance decrease, resulting in an overall conductance increase at higher frequencies.

Compiling the above results into a total conductance-frequency response curve yields the results shown in Fig.~\ref{fig:Fig2}(g), where the non-monotonic frequency dependence shows that our integrated iontronic circuit reproduces biological SRDP. Here the total conductance $g$ is normalised by the circuit's conductance $g_{\text{init}}$ during the first pulse. A positive conductance offset is visible in the experimental results (blue) compared to the theory (red), likely due to the aforementioned constant positive conductance offset for the short channels as shown in Fig.~\ref{fig:Fig2}(f,bottom). Barring this small offset, the frequency dependence corresponds to SRDP within both theory and experiment. At low frequencies a decrease in conductance emerges with increasing frequency, corresponding to the decrease in synaptic strength for lower frequencies as depicted in Fig.~\ref{fig:Fig2}(a). Then for higher frequencies the overall conductance increases, corresponding to the increase in synaptic strength at higher spike train frequencies. With this non-monotonic (and thus nonlinear) frequency response we can separate (real-world) data that is not linearly separable as we will show next.

\section{Audio anomaly detection with nonlinear frequency response}
We demonstrate the reliability and functionality of the non-monotonic frequency response SRDP, shown in Fig.~\ref{fig:Fig2}, with a simple analysis of real-world data. We applied the dual-channel circuit to classify sound recordings from three insect species, dragonflies, bees, and hornets. In particular, the aim is to separate the bee-generated sound signals from those of its predators, dragonflies and hornets. We used 20 audio signals of 0.4\,s per insect, which were first converted into voltage pulse trains using zero-crossing detection with bin sizes of 0.1\,s \cite{Chen2024NeuromorphicMemristor}. This procedure is illustrated in Fig.~\ref{fig:Fig3}(a), which shows one of the 20 audio signals for each of the three insect species (top) and its corresponding voltage pulse train (bottom). The three scatterplots (middle) of the number of zero-crossings show the 20 individual counts during each of the 0.1\,s intervals (dots), their average (blue rectangles), and their standard deviation (bars). The number of zero crossings during 0.1\,s intervals is converted into 0.1\,s spike train segments with a spike for each 4.4 crossings, resulting in characteristic frequencies of $\sim50$ Hz (dragonfly), $\sim192$ Hz (bee), and $\sim455$ Hz (hornet). Notably, for the binary classification of bee or predator, these inputs are not linearly separable based on these frequencies. Therefore, the signals have to be separated by a device that does not just respond differently to differing frequencies, but does so in a non-monotonic and nonlinear way. With the nonlinear SRDP, our dual channel fluidic circuit will serve as an artificial synapse that can perform this task, as schematically depicted in Fig.~\ref{fig:Fig3}(b).

When these signals were applied to the device, the measured current traces exhibited a clear dependence on insect type, as shown in Fig.~\ref{fig:Fig3}(c) where we depict various total circuit current measurements for different inputs corresponding to dragonflies (blue), bees (brown), and hornets (yellow). The low and high frequencies of dragonflies and hornets, respectively, increase the total circuit conductance and thereby generate stronger currents, whereas the intermediate bee frequencies yields a reduced current strength. Crucially, this response was reliable and reproducible. In Fig.~\ref{fig:Fig3}(d) we show the spread of the conductance at the end of the 0.4 s signal, determined by 20 measurements per insect, compared to the conductance after the first pulse $g_{\text{init}}$. Despite the limited differences in conductance of a few percent, the device is sufficiently reproducible and accurate that there is a significant and reliable separation between the inputs. Both low (dragonfly) and high (hornet) frequencies produced higher conductances compared to the intermediate frequency input from the bee. Therefore, the SRDP response nonlinearly separates the intermediate bee frequency from those of its predators, demonstrating the circuit's capability to act as a nonlinear kernel that transforms otherwise inseparable temporal patterns such that they become linearly discriminable, for example, by a standard (ionic) crossbar array \cite{Liu2025Resistance-RestorableChip,VanDeBurgt2018OrganicComputing,Kazemzadeh2025AllArray,Xu2025Angstrom-Scale-ChannelComputing,Zhang2022AdaptiveConductor,Hu2023AnComputing,VanDeBurgt2017AComputing}.

\section{Discussion}
We have presented an integrated fluidic circuit that exhibits biologically plausible synaptic Spike Rate Dependent Plasticity (SRDP) and demonstrated its use as a hardware-based nonlinear kernel to classify data that is not linearly separable. These features were predicted by a theoretical model and the essentials of the internal dynamics of this circuit are theoretically well-understood. This enabled a targeted experimental design to achieve SRDP, while some quantitative issues remain to be investigated in more detail. Moreover, the general concepts demonstrated here should transfer directly to other types of (iontronic) memristors \cite{Kamsma2024Brain-inspiredNanochannels,Zhang2024GeometricallySystems}.

To achieve SRDP, the circuit incorporates two different internal timescales as well as depression and potentiation features by varying channel lengths and orientations, respectively. Theoretical predictions of individual channel membranes were first experimentally verified, including a general device-agnostic prediction of the relation between the frequency of an oscillating applied voltage and the enclosed area inside the resulting current-voltage hysteresis loop. These device level theoretical understandings enabled straightforward by-design heterogeneous timescales and depression-potentiation characteristics, features that have been shown to provide a powerful framework for physical fluidic recurrent neural networks \cite{Kamsma2025EchoSubstrate}. This is of especial relevance as the power consumption of the individual channels could scale down to ultralow values \cite{Shi2023UltralowMemristor}, enabling highly energy-efficient iontronic networks \cite{Kamsma2026Energy-efficientCircuits}.

Combining these features in an integrated fluidic circuit resulted in biologically plausible SRDP, where a minimum circuit conductance is found at an intermediate frequency and a larger conductance at both lower and higher frequencies. This non-monotonic conductance response was able to reliably and repeatedly act as a nonlinear kernel for separating (encoded) real-world insect audio data that is otherwise not linearly separable, thereby opening possibilities for eventual linear classification via e.g.\ (fluidic) crossbar arrays \cite{Liu2025Resistance-RestorableChip}. 

The current results represent a relatively simple time series classification task within a specific frequency regime. Expanding the iontronic circuit to more integrated components, thereby enabling the classification of more complex signals across a wider range of frequencies, and coupling the output to a physical crossbar array \cite{Liu2025Resistance-RestorableChip} represent natural next steps. Moreover, coupling the input to a nanofluidic oscillator \cite{Xiong2025ANeuron} for input spike trains provides an interesting perspective for an end-to-end nanofluidic neuromorphic device, akin to earlier results in more conventional substrates \cite{Zolfagharinejad2025AnalogueComputing}. While the change in conductance is sufficient to reliably separate the signals, it is quantitatively a limited effect that could be improved in future work with iontronic devices that exhibit stronger current rectification, thereby widening the window of achievable conductances. Crucially, the principles demonstrated here remain applicable for such future iterations. Moreover, it has been shown that the inhibitory-excitatory response of the memristors can also be programmed via the concentration gradient over the devices \cite{Wang2025BioinspiredSynapses,Kamsma2025ChemicallyApplications,Portillo2024ReversibleMemristor,Portillo2026ReservoirMemristors}. This potentially offers a chemically regulated pathway for reconfiguring the SRDP response, akin to the chemical modulation by gamma-aminobutyric acid (GABA) receptors in biological neurons \cite{Xu2008GABABSynapses}. Similarly, the possibilities for multiple distinct physical input stimuli can be further expanded by leveraging iontronics' intrinsic pressure dependence \cite{Jubin2018DramaticNanopores,Barnaveli2024Pressure-GatedProcessing,Kamsma2025EchoSubstrate}.

In conclusion, we constructed, characterized, and integrated a fluidic iontronic device that is capable of analysing time series data. The underlying theory enabled a targeted fabrication to achieve its bio-inspired nonlinear Spike Rate-Dependent Plasticity and the fabrication demonstrates how these devices are integrated in one fluidic circuit to form a physical nonlinear frequency kernel that can reliably separate real-world data. Therefore, these results represent a step toward integrated fluidic (recurrent) neural network architectures and biologically plausible neuromorphic features.

\section{Materials and Methods}\label{sec:methods}
We fabricated conical nanochannels in poly(ethylene terephthalate) (PET) foils using standard asymmetric latent-track etching technique \cite{apel_diode-like_2001,siwy_fabrication_2002,zhou_nanofluidic_2024}. We irradiated 12-$\mu$m and 2.5-$\mu$m thick PET foils by swift heavy ions at the Lanzhou Heavy Ion Research Facility (HIRFL), creating latent tracks with areal densities of $3\times10^{8}\ \mathrm{cm^{-2}}$ and $5\times10^{9}\ \mathrm{cm^{-2}}$, respectively. We then exposed the membranes to ultraviolet light to promote photo-oxidation along the tracks. Next, we clamped the foils in a custom two-reservoir cell, filling one side with 9 M NaOH as etchant and the other side with a mixture of 1 M formic acid and 1 M NaCl as stopping solution. We monitored the ionic current at a constant 1 V bias to control the etching time and final pore size. After etching, we replaced both reservoirs with 0.1 M KCl and set exposed conductive areas of 78.5 mm$^{2}$ (12-$\mu$m foil) and 3.14~mm$^{2}$ (2.5-$\mu$m foil) to achieve comparable conductances. Finally, we inserted Ag/AgCl electrodes into the reservoirs and performed electrical characterization using an AMETEK ModuLab Xtreme workstation.

To characterize the morphology of the conical nanochannels, we performed transmission electron microscopy (TEM) after staining with ruthenium tetroxide (RuO$_4$) \cite{adla_characterization_2003,lu_angstrom-scale_2025}.
We first prepared $\sim$100-nm-thick cross-sectional slices (perpendicular to the membrane plane) using an ultramicrotome (Leica UC7/FC7) and exposed them to RuO$_4$ vapor for 20 min. RuO$_4$ reacts with functional groups on the inner channel walls, and the heavy-metal stain enhances TEM contrast to reveal the channel geometry.

The steady-state conductances used for the results in Figs.~\ref{fig:Fig1}(c) and (d) were calculated using the model of Ref.~\cite{Kamsma2023UnveilingIontronics}. We find that channels with base radii of 8 nm and 16 nm, and tip radii of 2 nm and 4 nm, for the long and short channels, respectively, reproduce the experimentally observed steady-state I-V data, shown in the insets of Figs.~\ref{fig:Fig1}(c) and (d). We note that, due to the variety of different channel geometries, these radii have a limited physical meaning and rather serve to reproduce the characteristic current rectification of the membranes. Besides this, the channel lengths and ion concentration match with the experimental values $L_{\mathrm{l}}=12$ $\mu$m and $L_{\mathrm{s}}=2.5$ $\mu$m and ion concentration of 100 mM, where the ions feature a diffusion coefficient of $D=1\text{ }\mu\text{m}^2\text{ms}^{-1}$. Lastly, a fixed surface charge of $e\sigma=-0.2\;e\text{nm}^{-2}$ is assumed, resulting in a typical surface potential of $\psi\approx-40$ mV. The steady-state currents that results from these parameters agree well with experiments, as shown in the insets of Figs.~\ref{fig:Fig1}(c) and (d). However, the membranes saturate to higher and lower conductances during the voltage pulse train experiments for the short and long channel membranes, respectively. Specifically, for the highest frequency of 455 Hz the short membrane saturates to a conductance 1.57 times higher than predicted, while the long channel membrane saturates to a conductance 0.97 times lower. Therefore, to account for this, the conductance contributions from the membranes are scaled by factors 1.57 and 0.97 for the short and long channels, respectively. This is only to enable keeping the parameters consistent with the ones used for calculations for Fig.~\ref{fig:Fig1}, for simplicity. The SRDP behaviour also emerges within theory without rescaling for other channel parameters.

\begin{acknowledgments}
Y.G., D.S., and Y.X.\ would like to acknowledge their funding NSFC 12388101, 12550001, 12241201 and the Innovation Capability Support Program of Shaanxi(Program No.2024RS-CXTD-15).
\end{acknowledgments}

%

\end{document}